\pdfoutput=1
\documentclass{article}
\usepackage{graphicx}
\usepackage{amssymb}
\usepackage{amsmath}
\usepackage{color}
\usepackage[ampersand]{easylist}
\usepackage{soul}
\usepackage{mycommands}
\usepackage{hyperref} 
\usepackage{comment}
\usepackage{authblk}
\usepackage[margin=1.25in]{geometry}

\title{Estimating the True Effect Size Distribution with SIMEX}
\author[1]{Zhaoqi Li} 
\author[2]{Daniel Ting}
\author[2]{Ilya Gorbachev}
\author[2]{Ehsan Emamjomeh-Zadeh}
\author[2]{Houssam Nassif}
\affil[1]{Stanford \\ \tt{zli9@stanford.edu}}
\affil[2]{Meta \\ {\tt\{dting, ilyagorbachev, ehsanez, 
houssamn\}@meta.com}}

\date{}

\begin{document}

\maketitle

\section{Introduction}
Large scale online experimentation inherently runs many experiments. 
Unfortunately, this also generates a multiple testing problem
and results in overstated gains that cannot be reproduced~\cite{Fiez2024Anduril}.
For experimenters who see their treatment underperform expectations after launch,
this can lead to mistrust in the system. 
Furthermore, experimentation platforms themselves must make a number of policy decisions
that depend on their ability to predict the performance of a policy.
For example, they may wish to choose a lower significance level for hypothesis tests
if the additional product launches yield higher overall gains
that exceed the cost of increased false positive rates or Type I error. 

While these issues can be addressed, for example by using Bayesian methods
to adjust overstated gains or by evaluating decision processes under a generative model,
these methods require estimating the distribution of true effect sizes
from noisy estimated effect sizes.
This paper's main contribution is a novel, non-parametric method
to estimate this based on the ideas of SIMulation-EXtrapolation (SIMEX)~\cite{cook1994simulation}.

Existing work in online experimentation~\cite{berman2022false, kohavi2024false}
examines the False Discovery Rate of experiments.
This problem is also not restricted to online experimentation
but is also encountered in the ``reproducibility crisis'' in academia
where {efforts like}~\cite{jager2014estimate, bartovs2022z, van2023new}
estimate additional properties of a corpus of experiments,
such as the power of tests for detecting true effect sizes
or the probability that experiments can be replicated. 

This problem of estimating the unobserved distribution of true effects
is related to the deconvolution problem. 
In the deconvolution problem, there are independent draws from a density $f$ and additive measurement error from $g$,
so that the observations are drawn from the convolution $h = f * g$.
One wishes to estimate the deconvolved density $f$. 
Several methods have been developed for this problem,
particularly in a line of work using  deconvolution kernel density estimators~%
\cite{stefanski1990deconvolving, delaigle2021deconvolution},
empirical Bayes normal means models (EBNM)~\cite{efron2016empirical, willwerscheid2021ebnm},
or simply normal mixture models. 
However, estimating effect size distributions in ABTest differ from most deconvolution problems
in two major ways. First, it allows for homoskedastic errors. Second, the signal-to-noise ratio is 
extremely high. The noise distribution is expected to have greater spread than the effect size distribution so that kernel methods are inappropriate for the problem.

More formally, {suppose} that $n$ experiments are run where
the unobserved true average treatment effect is denoted by $X_i$ for experiment $i$.
Further, {suppose} that the observed ATE estimate $\hat{X}_i$
is unbiased with known variance $\sigma_i^2$.
Let $\mathbb{F}_0$ denote the unobserved empirical distribution of the true effects $X_i$
and $\mathbb{F}_{\vec{\sigma}}$ denote the empirical distribution
of the estimated lifts $\hat{X}_i$. Our goal is to estimate $\mathbb{F}_0$. 
We can also consider the generative process where $X_i$ is drawn i.i.d.
from a distribution $F_0$.
That is
\begin{align}
    X_i &\sim F_0 \\ 
    \hat{X}_i | X_i &\sim Normal(X_i, \sigma_i^2).
\end{align}

Our goal then is to estimate the distribution $F_0$.

\section{Methodology}
Our goal is to learn the CDF of the underlying empirical effect distribution $\mathbb{F}_0$.
However, the true measurements $X_i$ of the effect size are never observed.
Instead, we observe noisy effect size measurements
$X_i + \epsilon_i \sim Normal(X_i, \sigma_i^2)$
and their biased empirical CDF $\mathbb{F}_1$.
We then wish to undo the effect of the noise process
that takes $\mathbb{F}_0 \to \mathbb{F}_1$.

We first describe the general idea behind bias correction using SIMEX.
SIMEX estimates the bias induced by the noise process by adding even more noise.
Suppose we were given data $\mathcal{X}$ without measurement error,
with some estimator or function $\phi(\mathbf{X})$ of interest.
We wish to obtain a good estimator even when the observations
$\mathcal{X} + \mathcal{E}$ contain measurement error $\mathcal{E}$. 
By adding even more noise, we can compute a random function 
$\theta(c) \stackrel{d}{=} \phi(\mathcal{X} + c \cdot \mathcal{E})$
for any $c \geq 1$. This constitutes the SIMulation component. 

It is easy to see that the desired estimate {is} $\theta(0) = \phi(\mathcal{X})$.
Thus, our goal is to Extrapolate the function $\theta(c)$
for $c \geq 1$ to the desired estimate $\theta(0)$. 
By fitting a parametric regression function $\hat{\theta}$
on the simulated values $\theta(c)$ for $c > 1$,
we can extrapolate to get our final estimate $\hat{\theta}(0)$. 
In summary, the general SIMEX procedure is:
\begin{enumerate}
    \item Simulate noise to compute
        $\theta(c) = \phi(\mathcal{X} + c \mathcal{E}')$ for $c \geq 1$.
    \item Fit a smooth function to $\theta(c)$, $c \geq 1$.
    \item Extrapolate to $\hat{\theta}(0)$.
\end{enumerate}

For our application in effect size estimation, we choose the function of interest $\phi_q$
to be the $q^{th}$ quantile. We estimate a grid of quantiles
to obtain an estimate of the inverse CDF $\mathbb{F}^{-1}$
as opposed to directly trying to estimate the CDF.
We found, however, that the typical quadratic regression used in SIMEX
was poor at extrapolation.
Furthermore, independently estimating each $\phi_q(\mathcal{X})$
would sometimes result in the inverse CDF being non-monotone.

This work introduces two improvements to address these drawbacks.
First, our choice to estimate quantiles rather than the CDF
allows us to construct a basis that yields both:
1) consistent estimates of the effect size distribution
if it belongs to a given parametric family,
and 2) good empirical estimates even when it is not in the parametric family.
Second, we extend the isotonic regression to ensure
all extrapolated estimates $\hat{F}^{-1}_c$ are monotonic  for all $c$.

\section{Experiments}

\begin{figure*}[ht]
    \centering
    \begin{minipage}{0.45\textwidth}
        \centering
        \includegraphics[scale=0.2]{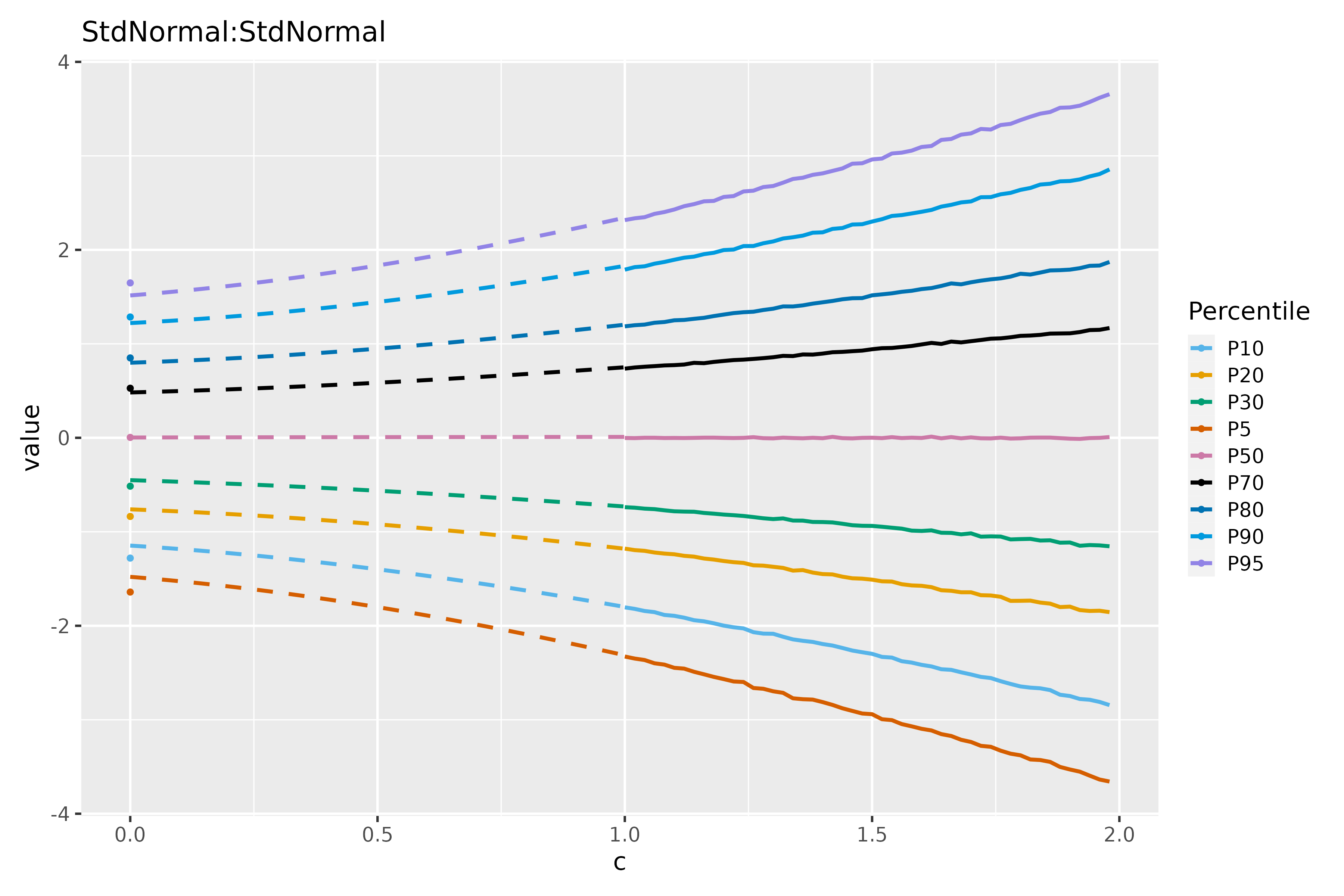}
        \caption{SIMEX curves for different percentiles}\label{fig:simex}
    \end{minipage}%
    \hfill 
    \begin{minipage}{0.45\textwidth}
        \centering
        \includegraphics[scale=0.2]{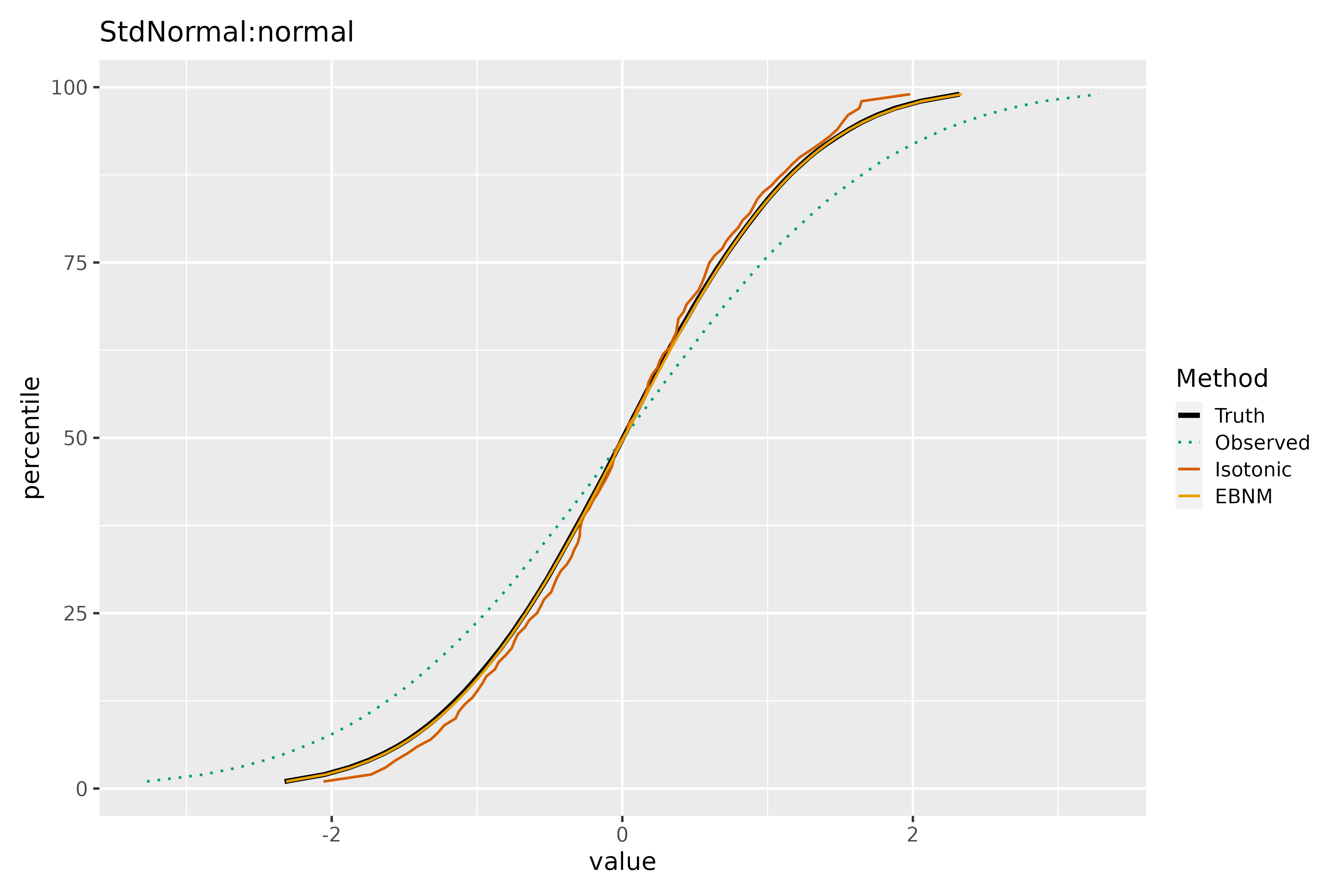}
        \caption{Estimated CDF for different methods}\label{fig:cdf}
    \end{minipage}    
\end{figure*}

We provide an experiment to illustrate our method.
We consider the simple case where the true effect distribution $\mathbb{F}_0=N(0,1)$,
and measurement error is also $N(0,1)$.
Figure~\ref{fig:simex} show in solid lines the evolution of each quantile $\theta(c)$
as more noise $c$ is added, while dotted lines show the estimated quantiles
as noise is removed through extrapolation.
Figure~\ref{fig:cdf} shows that we appropriately tighten
the observed distribution of estimated lifts and recover the true effect distribution.
We also compare it to EBNM~\cite{willwerscheid2021ebnm} which applies a parametric model, and our non-parametric method is nearly as good. While not shown here, our method also works for other effect size distributions. It also demonstrates some robustness properties as the quantiles underlying the method inherently enjoy some level of robustness.

\bibliographystyle{plain}
\bibliography{references}
\end{document}